\documentclass[pre,preprint,superscriptaddress]{revtex4-1}
\usepackage{subfigure}
\usepackage{graphicx}
\usepackage{epsfig}
\usepackage{amsmath}
\usepackage{epstopdf}
\usepackage{color}
\usepackage{mathtools}

\usepackage{hyphenat}
\let\oldref\ref
\renewcommand{\ref}[1]{(\oldref{#1})}
\begin{document}

\title{Statistical mechanics of a polymer chain attached to the interface of a 
cone-shaped channel} 
\author{Sanjay Kumar${^1}$, Sanjiv Kumar${^1}$, Debaprasad Giri${^2}$ and Shesh Nath${^1}$}
\affiliation{${^1}$Department of Physics, Banaras Hindu University, Varanasi 221005, India \\
${^2}$Department of Physics, Indian Institute of Technology(BHU), Varanasi 221005, India}


\begin{abstract}
A polymer chain confined in nano-scale geometry has been used to investigate
the underlying mechanism of Nuclear Pore Complex (NPC), where transport of 
cargoes is directional. It was shown here that depending on the solvent 
quality (good or poor) across the channel, a polymer chain can be either 
inside or outside the channel or both. Exact results based on the short 
chain revealed that a slight variation in the solvent quality can drag 
polymer chain inside the pore and {\it vice versa} similar to one seen 
in NPC. Furthermore, we also report the absence of crystalline (highly dense) 
state when the pore-size is less than the certain value, which may have 
potential application in  packaging of DNA inside the preformed viral proheads.
\end{abstract}

\maketitle
Understanding of equilibrium properties of biopolymers confined in a nano-scale geometry 
may delineate the possible mechanism involved in many biological processes {\it e.g.} 
translocation,  transport of proteins from the nucleus, ejection of viral DNA from the 
capsid, etc. 
\cite{degennes,Chakrabarti,vander1,muthu,Reisner, ralf,Craighead,hoog,Berndsen,Molineux,Marenduzzo}. 
Such processes have potential applications in designing nanotechnology devices including 
polymer separation, DNA sequencing, protein sensing etc. \cite{fu,wei,Wanunu,Venkatesan,marie,Branton,Nikoofard}.  
While most experiments and theories focus on driven systems \cite{Meller,Dekker,Henrickson,Meller1,Mathe,Sung,muthu1,Muthu2,Muthu3}, 
there are also considerable  
interests related to the unforced translocation \cite{Chuang,Luo, Huopaniemi,Panja,Kapahnke,Loerscher}. 
An interesting example is  when a pore connects two volumes of the solvent of different quality 
in case of nascent polypeptides, which translocates from the 
cytoplasm of eukaryotic cells to the lumen of the endoplasmic reticulum \cite{Rapoport}. There are few studies related 
to this phenomenon where scaling in translocation time has been reported \cite{Panja,Kapahnke,Loerscher}. 
However, besides the scaling in translocation time, understanding of the equilibrium properties of a polymer chain 
attached to the edge of a pore-interface of two liquids still remains elusive.

\begin{figure}
\begin{center}
\includegraphics[scale=0.3]{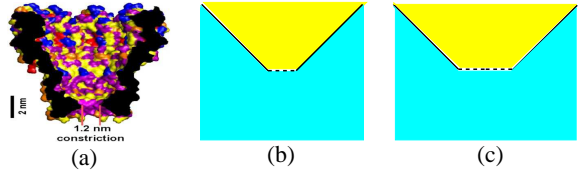}
\end{center}
\caption{Figure (a) shows the Crystal structure of MspA which looks like  cone shaped channel 
(taken from Ref. [35]).  Figures (b and c)  are the 
schematic representations of it for different pore-sizes ($r_p$). The thick lines represent
the impenetrable walls of the cone-shaped channel, which separate two volumes of the
liquid. The dashed line shows the penetrable interface of two liquids.
}
\label{fig-1}
\end{figure}

It is pertinent to mention here that in some cases the shape of the pore-interface  ({\it e.g.} Mycobacterium smegmatis
porin A (MspA), HIV-1 capsid), looks similar to the cone-shaped channel (Fig.1) \cite{Derrington,Zhao}. Though, the polymer 
translocation through the cone-shaped channel has been studied experimentally \cite{Lan,Sexton,Siwy}, the interest here 
is for the theoretical understanding in the framework of statistical mechanics \cite{ Halprin,Maghrebi,Nikoofard2,Hammer}.
Moreover, for the translocation of polymer from pore to outside, it was shown that the 
pore-size ($r_p$) and the size of polymer play an important role \cite{Bhattacharya,Hinterdorfer,Wanunu1,Ikonen}.
The radius of gyration, which gives the information about the size of the polymer chain,
scales as $N^\nu$, where $N$ is the number of monomers, and $\nu$ is  the gyration exponent. In the globule state 
(low-temperature), $\nu =1/d$, while at high-temperature polymer is in the swollen state,  and its value is given by 
the Flory approximation $\nu = \frac{3}{d+2}$ \cite{degennes,Chakrabarti,vander1}. Here, $d$ is the dimension. For a chain of 
finite length, high precession numerical simulations also show the existence of crystalline-like (highly-dense) 
state \cite{binder,Taylor,janke,kumpre} of polymer in addition to globule phase at low temperature. This is in 
accordance with findings of Doniach et al. \cite{doni}.  However, they termed it as ``molten globule" state. 
Interestingly, many viruses use molecular motors that generate large forces to package DNA to near-crystalline (high) densities 
inside preformed viral proheads \cite{Berndsen}. Therefore, it is prerequisite to explore the existence of such states in the 
confined geometry and its relationship with pore-size. 

\begin{figure}
\begin{center}
\includegraphics[scale=0.3]{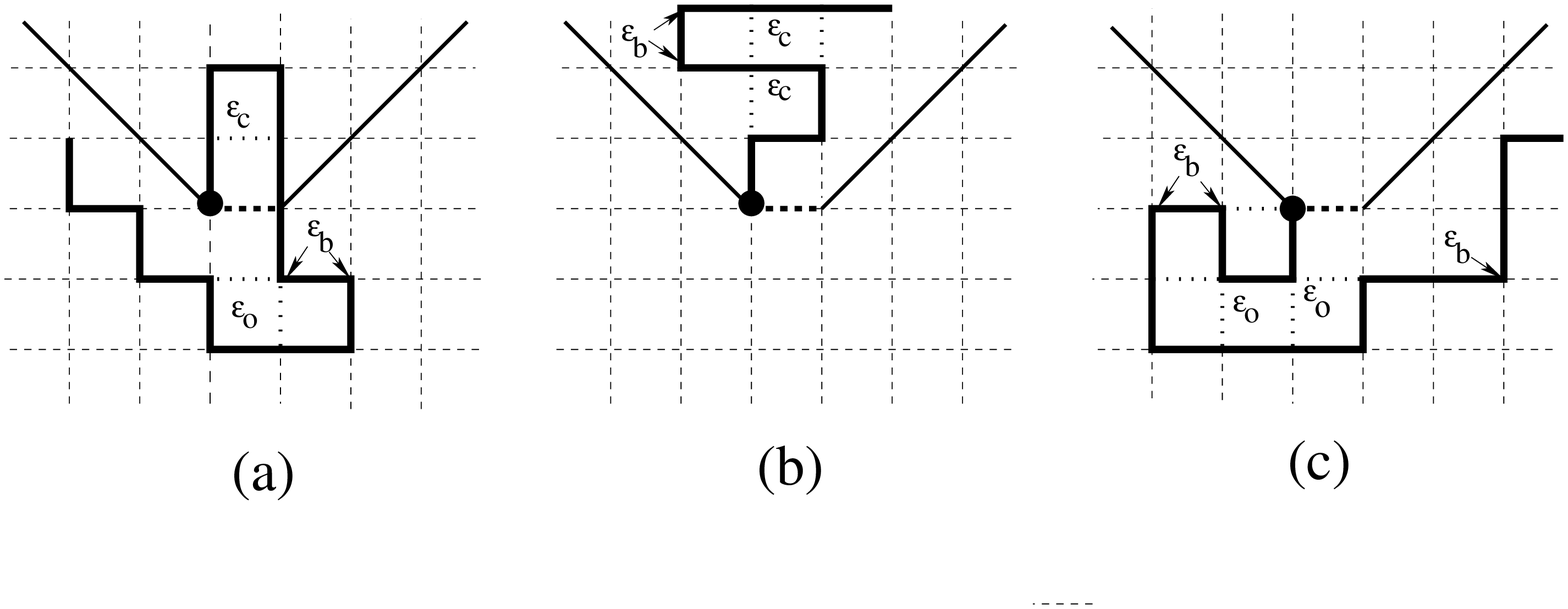}
\end{center}
\caption{Schematic representations of a polymer chain on the square
lattice having the cone-shaped channel of pore-size one.
$\epsilon_c$ and $\epsilon_o$ shown by the dotted 
lines correspond to the attractive interaction
between non-bonded monomers inside and outside the cone-shaped channel, respectively. 
$\epsilon_b$ represents the bending energy. 
Depending on the solvent quality, the polymer chain can stay both sides (Fig.2 a), or only
inside (Fig.2 b), or outside the cone-shaped channel (Fig.2 c).}
\label{fig-2}
\end{figure}

A cone-shaped channel of varying pore-size on the square lattice has been
constructed to model the nanopore. Two walls of the channel separate two liquids
in such a way that $\approx \frac{1}{4}$ volume is available to one type of liquid inside the 
pore (from the interface and above), while the remaining $\approx \frac{3}{4}$  volume is available to the 
liquid outside the pore (Fig.1). One end of the polymer chain is fixed at the interface of the
cone-shaped channel, whereas the other end is free to be anywhere (Fig.2). The polymer chain is not allowed to cross
the wall of the channel, except through the pore at the interface. The aim of the present study is two-fold: first to 
understand the effect 
of asymmetry arising due to the cone-shaped channel, and secondly to investigate the role 
of the solvent quality  on the equilibrium properties of a polymer confined in a pore. 
For this, we consider a self-attracting self-avoiding walk model of semi-flexible polymer,  
and use the exact enumeration technique \cite{degennes,Chakrabarti,vander1,kumpre} to obtain the equilibrium properties. 
Since, we have the exact information about the density of states, therefore, it is possible to explore the low-temperature 
behavior of the system and its dependence on the pore-size.
Previous studies for other systems have shown that the chain length
considered here is sufficient to predict the correct qualitative behavior, and  increasing the chain length
only yields a better estimate of the phase boundary \cite{Singh1,Singh2,Kumar1,Kumar2}.
The partition function of such a composite system  may be written as
 
\begin{figure}
\vspace {.25in}
\begin{center}
\includegraphics[scale=0.5]{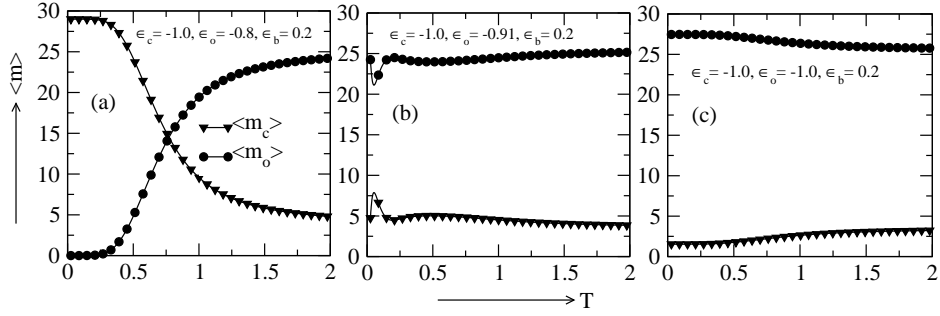}
\end{center}
\caption{Figures show the variation in average number of monomers ($m_c$ and $m_o$) 
with temperature for different values of $\epsilon_o$ at fixed $\epsilon_c$ and $\epsilon_b$. 
It is apparent from Fig. (b) that even when the solvent inside the pore is relatively 
poor than outside, polymer prefers to stay outside at all temperatures.
}
\label{fig-3}

\end{figure}
\begin{equation}
Z(T) = \sum_{(N_{pc}, N_{po}, N_b)} C_N (N_{pc}, N_{po}, N_b) u^{N_{pc}} \omega^{N_{po}} b^{N_b}.
\end{equation}

Here, $C_N (N_{pc}, N_{po}, N_b)$ is the total number of different conformations of walk of length $N=28$ steps (29 monomers). 
$N_{pc}$ and $N_{po}$ are the nearest-neighbor pairs inside  and outside the cone-shaped channel (Fig.2), respectively. Here, 
$N_b$ is the number of bends in the  chain. $\omega = \exp (-\beta \epsilon_o)$, $u = \exp (-\beta \epsilon_c)$, and 
$b = \exp (-\beta \epsilon_b)$, are the Boltzmann weights of nearest-neighbor interaction outside ($\epsilon_o$)  and 
inside ($\epsilon_c$) of the channel, and bending  energy ($\epsilon_b$) respectively. 
If $\epsilon_b \le 0$, $b \ge 1$, and this corresponds to a flexible chain.
The semiflexibility of the chain may be introduced by assigning $b < 1$ {\it i.e.}  $\epsilon_b \ge 0$.
$\beta = \frac{1}{k_B T}$, where $k_B$ is the Boltzmann  constant and $T$ is the temperature. In the following, 
we set $\frac{\epsilon_c}{k_B} = 1$ and do the analysis in the reduced unit. 
The value of the transition temperature
for  finite $N$ can be obtained from  the peak value of the  specific heat $ c = \frac{\chi_c}{T^2}$, where fluctuation ($\chi_c$) inside 
cone is defined as $\langle N_{pc}^2\rangle -\langle N_{pc} \rangle^2$, with the $k$th moment given by
\begin{equation}
\langle N_{pc}^k \rangle = \frac{1}{Z} \sum_{(N_{pc}, N_{po}, N_b)} N_{pc}^k C_N (N_{pc}, N_{po}, N_b ) u^{N_{pc}} \omega^{N_{po}} b^{N_b}.
\end{equation}
Similarly, one can calculate specific heat corresponding to non-bonded nearest-neighbor pairs outside the channel. The average number of monomers 
inside ($m_c$)  (outside  ($m_o$)) the pore is given by

\begin{equation}
\langle m_c \rangle =\frac{1}{Z}\sum_{(N_{pc}, N_{po},N_b, m_c, m_o)}\hspace{-0.3in}m_c C_N(N_{pc},N_{po},N_b, m_{c},m_{o}) 
u^{N_{pc}} \omega^{N_{po}} b^{N_b}.
\end{equation}

In Figs.3 (a-c), we have plotted the average number of monomers inside and outside the pore as a function
of temperature for different values of $\epsilon_o$ at fixed $\epsilon_c = -1$ and $\epsilon_b = 0.2$. 
If outside solvent is comparatively good in nature,
the polymer prefers to be inside the pore at the lower temperature. With increase in temperature, $m_c$ decreases 
and consequently, $m_o$ increases. As solvent quality outside becomes poorer, $m_o$ increases further, and at 
high temperature, polymer prefers to be outside (Fig.3 a).  

The most striking feature one observes, when $\epsilon_o$ approaches to $\epsilon_c$.  At low temperature, 
there is a sudden change and almost all monomers prefer to stay outside (in a good solvent) rather inside the
pore which has a relatively poor solvent (Fig.3 b). With increase in the temperature,
one can see a  tendency of decrease in $m_o$ (consequently increase in $m_c$), but at high 
temperature, polymer prefers to be outside the pore. When solvent quality on both sides
becomes similar, the tendency of decrease in $m_o$ also vanishes, and  polymer  remains outside of the
pore (Fig.3 c). If outside solvent becomes relatively poor than that of the inside, polymer 
prefers to stay outside. A similar behavior is observed, when $\epsilon_o = -1$ is kept fixed
and $\epsilon_c$ is varied \cite{sm}. In this case, polymer remains outside the pore.
Even if solvent inside the pore becomes poorer, polymer stays outside the pore.
To drag polymer inside the pore, one has to choose 
$\epsilon_c \leq -1.2$ at low temperature \cite{sm}. This is because of the competetion between 
the reduction of entropy arising due to the confinement and gain in the energy due to the increase in 
monomer-monomer attaction. Ejection of monomers from the cone depends on the difference of the free energy 
accross the interface, which either be overcome by change in solvent quality or temperature. 

The free energy barrier may be calculated from the principle of detailed balance of energy:
\begin{equation}
\frac{k_{c-o}}{k_{o-c}} = \frac{P_c}{P_o}
\end{equation}
where $k_{c-o}$ is the rate coefficient of ejection  from the cone side ($c$) to outside ($o$). 
The rate coefficient is assumed to follow the Arrhenius kinetics \cite{kinetic}, $k_{c-o} = 
k^{*} \exp[-\beta \Delta f]$. where $\Delta f$ denotes the free energy barrier 
associated with ejection from cone side to outside and $k^*$ is a constant. In Fig. 4, we plot
the $\ln (\frac{P_c}{P_o})$  with $\beta$, which is linear in nature. The slope of the curve 
gives the height of the free energy barrier, which is in good agreement with the free energy 
difference of polymer (all in and all out) obtained from the partition function for a given 
set of parameters. 

\begin{figure}
\vspace{.25in}
\begin{center}
\includegraphics[scale=0.4]{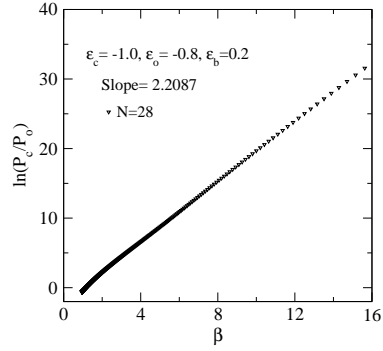}
\end{center}
\caption{Variation of $\ln (\frac{P_c}{P_o})$ with $\beta$. The slope gives the 
free energy barrier across the  interface ($c-o$).}
\label{fig-4}
\end{figure}

\begin{figure*}
\vspace{.25in}
\begin{center}
\includegraphics[scale=0.4]{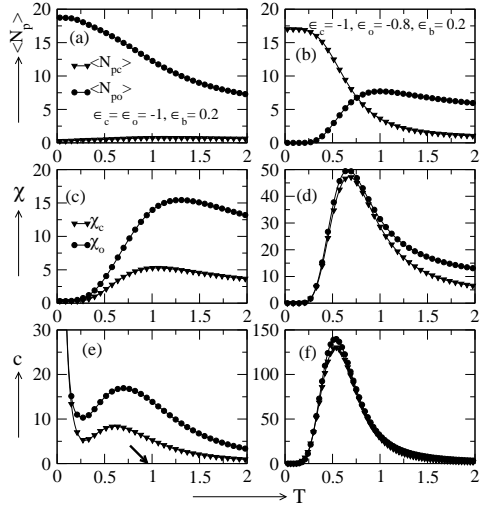}
\end{center}
\caption{Figures (a-b) show the variation of $\langle N_{po} \rangle$ and $\langle N_{pc} \rangle$ with
temperature for different sets of interactions at fixed $\epsilon_b$. Figs. (c-d)  and (e-f) show the variation 
of fluctuation and specific heat with temperature, respectively. The arrow on x-axis (Fig. e) corresponds 
to coil-globule transition in free space (in absence of cone shaped channel). 
}
\label{fig-5}
\end{figure*}

In Figs.5 (a-b), we plot the variation of  $\langle N_{pc}\rangle$ and 
$\langle N_{po}\rangle$ with temperature for different sets of interactions at fixed $\epsilon_b$.  
If the nature of the solvent on both sides is the same ($\epsilon_c = \epsilon_o = -1$), 
$\langle N_{po} \rangle$ decreases with temperature, whereas $\langle N_{pc} \rangle$ remains almost negligible. 
The system undergoes globule-coil transition outside the pore, which can be seen from the fluctuation curve 
(Fig.5 c) or the specific heat plot (Fig.5 e).  It may be noted that the transition temperature 
of polymer chain in free space is $\sim 0.93$ \cite{kumpre} shown by arrow  on x-axis in Fig. 5 e.  This shift 
is because of reduction of entropy induced by the cone shaped channel. 
For a relatively poor solvent inside the pore ($\epsilon_c = -1 $ and $\epsilon_o = -0.8$, and $\epsilon_b = 0.2$), 
we find that $\langle N_{pc} \rangle$ decreases with temperature much faster (Fig.5 b). 
Instead of going to the
coil state in the pore, increase in $\langle N_{po} \rangle$ indicates that the polymer prefers to be in the globule state 
outside the pore. With further increase in temperature,
polymer acquires coil conformations outside the pore (Figs.5 d $\&$ f). When the outside solvent is relatively 
poor than that of the inside, $\langle N_{po} \rangle$  decreases with temperature, and globule-coil transition 
occurs outside the pore.

In order to rule out that this is not an artifact of the lattice model or a finite-size effect, we 
revisited the model with a flat interface of a pore-size one. One end of the polymer chain is attached to the
edge of the pore. In this case, the polymer will not experience any confinement, but the interface will 
separate two volumes of the liquid. We show the variation in the average number of monomers on both sides of the 
solvent with temperature for four different sets of interactions (Fig.6). In the 
absence of confinement, a major fraction of polymer chain
prefers to stay in the poor solvent side at low temperature. At high temperature or when both sides of the solvent are the same,
there is no preferential choice,
and monomers are uniformly distributed {\it i.e.} half of the monomers stays on one side and the remaining half to the other side.
These results provide unequivocal support that confinement arising due to the cone-shaped channel gives rise
to such an effect, and we anticipate that experiments will be able to verify these findings.

\begin{figure}
\vspace{.1in}
\begin{center}
\includegraphics[scale=0.3]{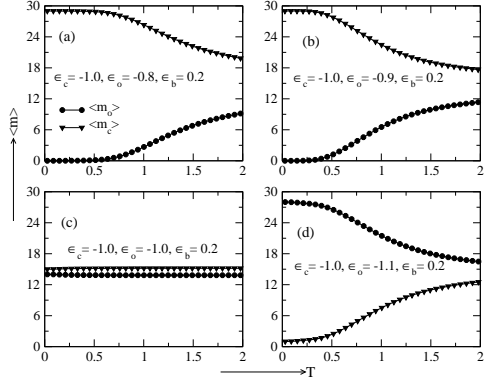}
\end{center}
\caption{Same as Figs.3, but for the flat interface of pore-size one. Here, $m_c$ corresponds to
the number of monomers on one side of the solvent, whereas $m_o$ for the other side. Fig. (d) shows 
to drag polymer in the preferred  side, one has to increase the monomer-monomer attraction of that side.}
\label{fig-6}
\end{figure}

For $N= 28$, the average size of the polymer is $\approx 5$ and $\approx 12$  at low and high temperature, respectively. 
If $r_p$ is less than 5, migration of polymer from inside to outside at low temperature should be difficult. For this, a polymer has to 
first unfold and then cross the interface, therefore, dynamics appears to be slow. Whereas for a bigger pore-size (5 and above), dynamics
would be fast. In view of above, we enumerate all possible walks of step $28$ for different $r_p$ (Fig.1), 
and calculate the partition function for each case. Here, we focus now only on the solvent quality 
inside  the pore, which is more poor than that of the outside {\it i.e.} $\epsilon_c = -1$ and $\epsilon_o = -0.8$ (Fig.3 a).
 
\begin{figure}[b]
\vspace{.22in}
\begin{center}
\includegraphics[scale=0.3]{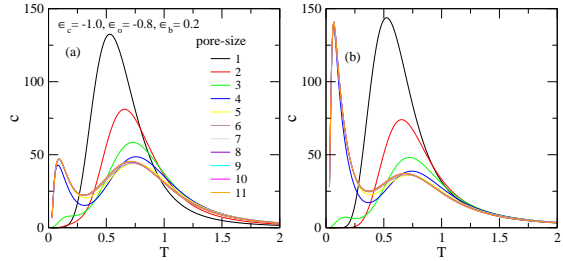}
\end{center}
\vspace{.1in}
\caption{Variation of specific heat with temperature for different pore-sizes for a given set of interactions:
(a) inside the pore; (b) outside the pore. Appearance of the second peak corresponds to the high-dense state,
which is apparent for the pore-size more than four.}
\label{fig-7}
\end{figure}

In Fig.7 a, we depicted the variation of specific heat $c$ inside the pore with temperature
for different pore-sizes ($r_p = 1-11$). One can see from the plot that when $r_p$ is less than 4, the
system has one peak corresponding to globule-coil transition and the transition temperature shifts 
to the right as pore-size increases. Above the pore-size 4, the system exhibits two peaks. 
This is in accordance with earlier studies which exhibited frozen structure (beta sheet) for the semiflexible chain at 
low temperature \cite{binder,Taylor,janke,kumpre,doni}. 
In Fig.8, we plot the density of states $D(N_p)$ as a function of nearest-neighbor contacts ($N_p$) at different temperature 
(around the peak positions). 
It is evident from the plot that at high temperature, polymer is in the coil-state, where contributions are from $N_p=1-19$.
Near the globule-coil transition, contributions are from $N_p=11-19$. This implies that globule has some voids, and therefore, 
entropy of the globule state is relatively high. At low temperature, the dominant contribution is from $N_p =19$ only, 
which corresponds to the highly-dense state, where all lattice sites are occupied.  
Thus, emergence of the new peak is the signature of high-dense state (frozen) to globule transition \cite{binder,janke}, whereas 
the second peak corresponds to globule-coil transition.  For the pore-size 6 and higher, all plots fall  on each other indicating 
that pore-size has no effect for 
a given length. Similar behavior has been seen for the polymer chain outside the pore (Fig.7 b).
 
\begin{figure}
\begin{center}
\includegraphics[scale=0.3]{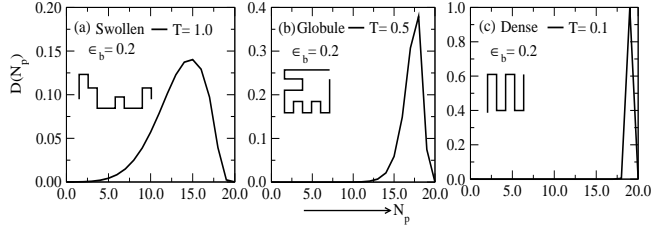}
\end{center}
\caption{Density of states as a function of $N_p$ for different temperatures at fixed $\epsilon_b$. At low temperature,
all sites are occupied whereas globule has some voids.}
\label{fig-8}
\end{figure}

To the best of our knowledge, this is the first study of correlating the effect of temperature, solvent quality, and pore-size 
on a polymer chain attached to the edge of a cone-shaped channel. When outside solvent is poorer than the inside,
polymer always stays outside the pore. This is because the free energy of the system is less
compared to the inside. Even if the solvent quality inside and outside channel is the same, polymer prefers to
stay outside. This is due to the reduction of entropy inside the pore than the outside, which pushes
polymer from the pore. The most surprising finding is the ejection of polymer from the pore and 
{\it vice versa} with a slight change in solvent quality. This happens in the case
when the solvent quality inside is relatively poor than the outside. The competition between gain in energy
and loss of entropy  inside the pore, together with a gain of entropy outside the pore leads to this interesting
behavior.  Such process one finds in case of nuclear pore complex (NPC), where transport through the NPC is directional 
such that many cargoes are only imported into or exported from the nucleus, although other cargoes do shuttle in and out 
continuously \cite{din1}. 
The estimation of barrier height across the interface (Fig. 4) suggests that polymer of different 
length may collapse on a single curve. In fact for a small chain we do find such collapse \cite{sm}, however, to 
be sure one has to go for higher chain length preferably using high precesion Monte Carlo simulation. 

The absence of crystalline (highly-dense) state, when the pore-size is less than four, may be 
understood recalling that
the average size of the polymer is $\approx 5$ for a given length. On a square lattice, one expects square
like frozen structure. Since, the polymer is attached to the edge of the pore, such structure cannot
exist in the cone-shaped channel. When the size of the pore approaches 5, one can see the signature of the first peak
corresponding to the high-dense state. Once the pore size exceeds 5, a polymer of present length will not
experience confinement and such structure can exist. One may recall that for the semi-flexible polymer chain ({\it e.g.} proteins), 
crystalline state is more prominent in the form of $\beta-$sheet at low temperature \cite{Kumar1}. Hence, at this stage our studies 
warrant further
investigation on the relation between the pore-size and the length of the polymer preferably using numerical simulations of a longer chain.

\acknowledgments
We thank Y. Singh for many  fruitful  discussions  on  the  subject. Financial assistance
from the Department of Science and Technology, India and   University Grants  Commission,  India  
are gratefully acknowledged.   One of  us (SK) would  like  to  acknowledge  ICTP, Italy, where a
part of the work has been carried out.

{}


\begin{thebibliography}{}
\bibitem{degennes}  P. G. de Gennes, {\it Scaling Concepts in Polymer Physics} (Cornell University Press, Ithaca, 1979).
\bibitem{Chakrabarti} B. K. Chakrabarti, {\it Statistics of Linear Polymers in Disordered Media} (Elsevier, Amsterdam 2005).
\bibitem{vander1} C. Vanderzande, {\it Lattice models of polymers} (Cambridge University Press: Cambridge, 1998).
\bibitem{muthu} M. Muthukumar, {\it Polymer Tanslocation} (CRC Press, 2011).
\bibitem{Reisner} W. Reisner, J. N. Pedersen, and R. H. Austin Rep. Prog. Phys. {\bf 75}, 106601 (2012).
\bibitem{ralf} V. V. Palyulin, T.  Ala-Nissilab, and R. Metzler, Soft Matter {\bf 10}, 9016 (2014).
\bibitem{Craighead} H. Craighead, Nature {\bf 442}, 387 (2006).
\bibitem{hoog} D. P. Hoogerheide, B. Lu, and J. A. Golovchenko, ACS Nano {\bf 8}, 7384 (2014).
\bibitem{Berndsen}Z. T. Berndsen, N. Keller, S. Grimes, P. J. Jardine, and D. E. Smith, PNAS, {\bf 111}, 8345, (2014) .
\bibitem{Molineux} I. J. Molineux and D. Panja, Nat. Rev. Microbio. {\bf 11}, 194, (2013).
\bibitem{Marenduzzo} D. Marenduzzo, C. Micheletti, E. Orlandini, and D. W. Sumners, PNAS {\bf 110}, 20081, (2013).
\bibitem{fu} J. Fu, R. B. Schoch, A. L. Stevens, S. R. Tannenbaum, and J. Han, Nat. Nanotech. {\bf 2}, 121 (2007).
\bibitem{wei} R. Wei, V. Gatterdam, R. Wieneke, R. Tampe, and U. Rant, Nat. Nanotech. {\bf 7}, 257, (2012).
\bibitem{Wanunu} M. Wanunu, Phys. Life. Rev. {\bf 9}, 125 (2012).
\bibitem{Venkatesan} B. M. Venkatesan and R. Bashir, Nat. Nanotech. {\bf 6}, 615 (2011).
\bibitem{marie} R. Marie {\it et al.}, PNAS {\bf 110}, 4893, (2013).
\bibitem{Branton} D. Branton et al., Nat. Biotechnol. {\bf 26}, 1146 (2008).
\bibitem{Nikoofard}   N. Nikoofard and H. Fazli, Phys. Rev. E {\bf 85}, 021804 (2012).
\bibitem{Meller}   A. Meller {\it et al.} PNAS {\bf 97}, 1079 (2000).
\bibitem{Dekker}   C. Dekker, Nat. Nanotech. {\bf 2}, 209 (2007).
\bibitem{Henrickson}   S. E. Henrickson, M. Misakian, B. Robertson, and J. J. Kasianowicz, Phys. Rev. Lett. {\bf 85}, 3057 (2000).
\bibitem{Meller1}   A. Meller, L. Nivon, and D. Branton, Phys. Rev. Lett. {\bf 86}, 3435 (2001).
\bibitem{Mathe}   J. Mathe {it et al.}, PNAS {\bf 102}, 12377 (2005).
\bibitem{Sung}  W. Sung and P. J. Park, Phys. Rev. Lett. {\bf 77}, 783 (1996).
\bibitem{muthu1}  M. Muthukumar, J. Chem. Phys. {\bf 111}, 10371 (1999); {\it ibid} {\bf 118}, 5174 (2003).
\bibitem{Muthu2}  M. Muthukumar, Phys. Rev. Lett. {\bf 86}, 3188 (2001). 
\bibitem{Muthu3}  M. Muthukumar and C. Y. Kong, PNAS {\bf 103}, 5273 (2006).
\bibitem{Chuang} J. Chuang, Y. Kantor, and M. Kardar, Phys. Rev. E {\bf 65}, 011802 (2001).
\bibitem{Luo} K. Luo, T. Ala-Nissila, and S.-C. Ying, J. Chem. Phys. {\bf 124}, 034714 (2006).
\bibitem{Huopaniemi} I. Huopaniemi, K. Luo, T. Ala-Nissila, and S.-C. Ying, J. Chem. Phys. {\bf 125}, 124901 (2006).
\bibitem{Panja} D. Panja, G. T. Barkema, and R. C. Ball, J. Phys.: Cond. Matt. {\bf 19}, 432202 (2007).
\bibitem{Kapahnke}   F. Kapahnke, U. Schmidt, D. W. Heermann, and M. Weiss  J. Chem. Phys. {\bf 132}, 164904 (2010).
\bibitem{Loerscher} C. Loerscher, T. Ala-Nissila, and A. Bhattacharya, Phys. Rev. E {\bf 83}, 011914 (2011).
\bibitem{Rapoport} T. A. Rapoport, Nature {\bf 450}, 663 (2007).
\bibitem{Derrington} I. M. Derrington, {\it et al.}, PNAS {\bf 107}, 16060 (2010).
\bibitem{Zhao} G. Zhao {\it et al.}, Nature {\bf 497}, 643 (2013).  
\bibitem{Lan}  W. J. Lan, D. A. Holden, B. Zhang, and H. S. White, Anal. Chem. {\bf 83}, 3840 (2011).
\bibitem{Sexton} L. T. Sexton, L. P. Horne, and C. R. Martin, Mol. Biosyst. {\bf 3}, 667 (2007).
\bibitem{Siwy} Z. Siwy and A. Fulinski, Phys. Rev. Lett. {\bf 89}, 198103 (2002).
\bibitem{Halprin} A. Halperin, Journal de Physique, {\bf 47}, 447 (1986).
\bibitem{Maghrebi} M. F.,Maghrebi, Y. Kantor, and M. Kardar Phys. Rev. E {\bf 86}, 61801 (2012).
\bibitem{Nikoofard2}  N. Nikoofard, H. Khalilian, and H. Fazli J. Chem. Phys. {\bf 139}, 074901 (2013)
\bibitem{Hammer} Y. Hammer and Y. Kantor,  Phys. Rev. E {\bf 92}, 062602 (2015).
\bibitem{Bhattacharya}   A. Bhattacharya {\it et al.}, Eur. J. Phys. E {\bf 29}, 423 (2009).
\bibitem{Hinterdorfer} P. Hinterdorfer and A. van Oijen, {\it Handbook of Single-Molecule Biophysics} (Springer, 2009).
\bibitem{Wanunu1} M. Wanunu, J. Sutin, B. McNally, A. Chow, and A. Meller, Biophys J. {\bf 95}, 4716 (2008).
\bibitem{Ikonen} T. Ikonen, A. Bhattacharya, T. Ala-Nissila, and W. Sung, J. Chem. Phys. {\bf 137}, 085101 (2012).
\bibitem{binder} K. Binder and W. Paul, Macromolecules {\bf 41}, 4537 (2008).
\bibitem{Taylor} M. P. Taylor, P. P. Aung, and W. Paul, Phys. Rev. E {\bf 88}, 012604 (2013).
\bibitem{janke} W. Janke and W. Paul, Soft Matter {\bf 12}, 642 (2016).
\bibitem{kumpre} S. Kumar and D. Giri, Phys. Rev. E {\bf 72}, 052901 (2005).
\bibitem{doni} S. Doniach, T. Garel, and H. Orland, J. Chem. Phys.{\bf 105}, 1601 (1996).
\bibitem{Singh1} Y. Singh, S. Kumar, and D. Giri, J. Phys. A: Math. Gen. {\bf 32}, L407 (1999).
\bibitem{Singh2} Y. Singh, D. Giri, and S. Kumar, J. Phys. A: Math  Gen. {\bf 34}, L67 (2001).
\bibitem{Kumar1}  S. Kumar and D. Giri, Phys. Rev. Lett. {\bf 98}, 048101 (2007).
\bibitem{Kumar2} S. Kumar {\it et al} Phys. Rev. Lett. {\bf 97}, 12801 (2006).
\bibitem{sm} S. Kumar {\it et al.} to be published.
\bibitem{din1} D. Osmanovic {it et al.} Soft Matter {\bf 9}, 10442 (2013).
\end{thebibliography}
\end{document}